\documentclass[prl,aps,twocolumn,superscriptaddress]{revtex4}
\usepackage{graphicx,psfrag}
\usepackage{amsmath,amssymb}
\usepackage[3D]{movie15}
\usepackage{hyperref}

\usepackage[utf8]{inputenc}

\newcommand\ve[1]{\mathbf{#1}}

\begin{document}
\title{Finding the ciliary beating pattern with optimal efficiency}
\author{Natan Osterman} 
\affiliation{J. Stefan Institute, Jamova 39, 1000  Ljubljana, Slovenia} 
\affiliation{Department of Physics, Ludwig-Maximilians
  University Munich, Amalienstrasse 54, 80799 Munich, Germany}
\author{Andrej
Vilfan}\affiliation{J. Stefan Institute, Jamova 39, 1000  Ljubljana, Slovenia} 
\affiliation{Faculty of Mathematics and Physics, University of
  Ljubljana, Jadranska 29, 1000 Ljubljana, Slovenia}
\email{andrej.vilfan@ijs.si}

\begin{abstract}
  We introduce a measure for energetic efficiency of biological cilia acting
  individually or collectively and numerically determine the optimal beating
  patterns according to this criterion. Maximizing the efficiency of a single
  cilium leads to curly, often symmetric and somewhat counterintuitive
  patterns. But when looking at a densely ciliated surface, the optimal
  patterns become remarkably similar to what is observed in microorganisms
  like Paramecium. The optimal beating pattern then consists of a fast
  effective stroke and a slow sweeping recovery stroke. Metachronal
  coordination is essential for efficient pumping and the highest efficiency
  is achieved with antiplectic waves.  Efficiency also increases with an
  increasing density of cilia up to the point where crowding becomes a
  problem.  We finally relate the pumping efficiency of cilia to the swimming
  efficiency of a spherical microorganism and show that the experimentally
  estimated efficiency of Paramecium is surprisingly close to the
  theoretically possible optimum.
\end{abstract}
 \maketitle 

Many biological systems have evolved to work with a very high energetic
efficiency. For example, muscle can convert the free energy of ATP hydrolysis
to mechanical work with more than 50\% efficiency
\cite{Kushmerick.Davies1969}, the F1-F0 ATP synthase converts electrochemical
energy of protons to chemical energy stored in ATP molecules with even higher
efficiency \cite{Yoshida.Hisabori2001}, etc.  At first glance, the beating of
cilia and flagella does not fall into the category of processes with such a
high efficiency. Cilia are hair-like protrusions that beat in an asymmetric
fashion in order to pump the fluid in the direction of their effective stroke
\cite{Sleigh74}. They propel certain protozoa, such as \textit{Paramecium},
and also fulfill a number of functions in mammals, including mucous clearance
from airways, L-R asymmetry determination and transport of an egg cell in
Fallopian tubes. Lighthill \cite{Lighthill1952} defines the efficiency of a
swimming microorganism as the power that would be needed to drag an object of
the same size with the same speed through viscous fluid, divided by the
actually dissipated power. Although the efficiency defined in this way could
theoretically even exceed 100\% \cite{Michelin.Lauga2010}, the actual swimming
efficiencies are of the order of $1\%$
\cite{Purcell1997,Chattopadhyay.Wu2006}.  In his legendary paper on life at
low Reynolds number \cite{Purcell1977} Purcell stated that swimming
microorganisms have a poor efficiency, but that the energy expenditure for
swimming is so small that it is of no relevance for them (he uses the analogy
of ``driving a Datsun [a fuel-efficient car of the period] in Saudi Arabia'').
Nevertheless, later studies show that swimming efficiency is important in
microorganisms.  In \textit{Paramecium}, more than half of the total energy
consumption is needed for ciliary propulsion \cite{Katsu-Kimura.Mogami2009}.

When applied to ciliary propulsion, Lighthill's efficiency
\cite{Lighthill1952} has some drawbacks. For one, it is not a direct criterion
for the hydrodynamic efficiency of cilia as it also depends on the size and
shape of the whole swimmer. Besides that it is, naturally, only applicable for
swimmers and not for other systems involving ciliary fluid transport with a
variety of functions, like L-R asymmetry determination
\cite{Supatto.Vermot2008}.  We therefore propose a different criterion for
efficiency at the level of a single cilium or a carpet of cilia. A first
thought might be to define it as the volume flow rate of the transported
fluid, divided by the dissipated power. However, as the flow rate scales
linearly with the velocity, but the dissipation quadratically, this criterion
would yield the highest efficiency for infinitesimally slow cilia, just like
optimizing the fuel consumption of a road vehicle alone might lead to fitting
it with an infinitesimally weak engine. Instead, like engineers try to
optimize the fuel consumption at a given speed, the well-posed question is
which beating pattern of a cilium will achieve a certain flow rate with the
smallest possible dissipation.

The problem of finding the optimal strokes of hypothetical microswimmers has
drawn a lot of attention in recent years. Problems that have been solved
include the optimal stroke pattern of Purcell's three link swimmer
\cite{Tam.Hosoi2007}, an ideal elastic flagellum \cite{Spagnolie.Lauga2010}, a
shape-changing body \cite{Avron.Kenneth2004}, a two- and a three-sphere
swimmer \cite{Alouges.Lefebvre2009} and a spherical squirmer
\cite{Michelin.Lauga2010}. Most recently, Tam and Hosoi optimized the stroke
patterns of \textit{Chlamydomonas}' flagella \cite{Tam.Hosoi2011}. But all
these studies are still far from the complexity of a ciliary beat with an
arbitrary three-dimensional shape, let alone from an infinite field of
interacting cilia. In addition, they were all performed for the swimming
efficiency of the whole microorganism, while our goal is to optimize the
pumping efficiency at the level of a single cilium, which can be applicable to
a much greater variety of ciliary systems.

So we propose a cilium embedded in an infinite plane (at $z=0$) and pumping
fluid in the direction of the positive $x$-axis. We define the volume flow
rate $Q$ as the average flux through a half-plane perpendicular to the
direction of pumping \cite{Smith.Gaffney2008}.  With $P$ we denote the average
power with which the cilium acts on the fluid, which is identical to the total
dissipated power in the fluid filled half-space. We then define the efficiency
in a way that is independent of the beating frequency $\omega$ as
\begin{equation}
  \label{eq:eff}
  \epsilon=\frac{Q^2}{P}\;.
\end{equation}
As we show in Appendix 1, minimizing the dissipated
power $P$ for a constant volume flow rate $Q$ is equivalent to maximizing
$\epsilon$ at a constant frequency. A similar argument for swimming efficiency
has already been brought forward by Avron et al. \cite{Avron.Kenneth2004}.

Furthermore, a general consequence of low Reynolds number hydrodynamics is
that the volume flow only depends on the shape of the stroke and on the
frequency, but not on the actual time dependence of the motion within a
cycle. This is the basis of Purcell's scallop theorem \cite{Purcell1977}. As a
consequence, the optimum stroke always has a dissipation rate constant in
time. We show this in Appendix 2.

We can make the efficiency $\epsilon$ completely dimensionless if we factor
out the effects of the ciliary length $L$, the beating frequency $\omega$ and
the fluid viscosity $\eta$. The velocity with which a point on the cilium
moves scales with $\omega L$ and the linear force density (force per unit
length) with $\eta \omega L$. The total dissipated power $P$, obtained by
integration of the product of the velocity and linear force density over the
length, then scales with $\eta \omega^2 L^3$. The volume flow rate $Q$ scales
with $\omega L^3$. Finally, the efficiency $\epsilon$ scales with $L^3/\eta$.
The dimensionless efficiency can therefore be defined as
\begin{equation}
  \label{eq:eff-dl}
\epsilon'= L^{-3}\eta \epsilon \;.
\end{equation}

When optimizing the efficiency of ciliary carpets, we have to use the measures
of volume flow and dissipation per unit area, rather than per cilium. We
introduce the surface density of cilia $\rho$, which is $1/d^2$ on a square
lattice. In the following we show that the volume flow generated per unit
area, $\rho Q$, is also equivalent to the flow velocity above the ciliary
layer.  The fluid velocity above an infinite ciliated surface namely becomes
homogeneous at a distance sufficiently larger than the ciliary length and
metachronal wavelength. The far field of the flow induced by a single cilium
located at the origin and pumping fluid in direction of the $x$ axis has the
form \cite{vilfan2006a}
\begin{equation}
  \label{eq:idealpump}
  \ve v(x,y,z)={\cal A} \frac {xz}{r^4} \hat e_r
\end{equation}
with an arbitrary amplitude $\cal A$. For this field the volume flow rate is
\begin{equation}
  \label{eq:volflowcalc}
  Q=\int_{-\infty}^\infty dy \int_0^\infty dz\, v_x (x,y,z)=\frac 2 3 {\cal A}
\end{equation}
and the velocity above an infinite field of such cilia is
\begin{equation}
  \label{eq:vinfty}
  v_c=\int_{-\infty}^\infty dx \int_{-\infty}^\infty dy \rho
  v_x(-x,-y,z)= \frac {2\pi}{3} \rho {\cal A} = \pi \rho Q\;,
\end{equation}
which is independent of $z$. In this regime, one can simplify the description
of cilia by replacing them with a surface slip term with velocity $v_c$
\cite{Julicher.Prost2009}.

We now define the collective efficiency as $\epsilon_c=(\rho Q)^2/(\rho P)$
and in dimensionless form as
\begin{equation}
  \label{eq:eff-coll}
  \epsilon_c'=\frac\eta L \frac { \rho^2 Q^2}{\rho P}\;.
\end{equation}
$\epsilon_c'$ is a function of the beat shape, the dimensionless density $\rho
L^2$ and the metachronal coordination, which will be explained later.
Additionally, for a single cilium or for collective cilia the efficiency also
depends on the dimensionless radius of the cilium, $a/L$, but this dependence
is rather weak, of logarithmic order.

At this point we note that our definition of efficiency is different from that
used by Gueron and Levit-Gurevich \cite{Gueron.Levit-Gurevich1999}.  They
define efficiency as volume flux through a specifically chosen rectangle above
the group of cilia divided by the dissipated power.  While this measure is
useful for studying the effect of coupling and metachronal coordination (they
show that the collective efficiency of a group of cilia increases with its
size), it lacks the scale invariance discussed above. Gauger et
al. \cite{Gauger.Stark2009} studied a model for individual and collective
magnetically driven artificial cilia. Rather than introducing a single measure
for the efficiency, they studied the pumping performance (which is the more
relevant quantity in artificial systems) and dissipation separately.  They
showed that the pumping performance per cilium can be improved with the proper
choice of the metachronal wave vector, while the dissipation per cilium
remains largely constant.  Both studies were limited to two-dimensional
geometry (planar cilia arranged in a linear row)
and neither of them uses a scale-invariant efficiency criterion proposed
here. On the other hand, Lighthill's criterion for swimming organisms shares
the same scaling properties as ours (it scales with the square of the swimming
velocity, divided by dissipation), but differs in definition because it
measures the swimming and not the pumping efficiency. At the end we will show
how the two measures are related to each other for a spherical swimmer.

Our goal is to find the beating patterns that have the highest possible
efficiency for a single cilium, as well as the beating pattern, combined with
the density and the wave vector that give the highest efficiency of a ciliated
surface.

\section{The model}

We describe the cilium as a chain of $N$ touching beads with radii $a$. The
first bead of a cilium has the center position $\ve x_1=(0,0,a)$, and each
next bead in the chain is located at $\ve x_{i+1}=\ve x_i + 2a(\sin\theta_i
\cos\phi_i, \sin\theta_i \sin \phi_i, \cos\theta_i)$. The maximum curvature of
the cilium is limited by the condition
\begin{equation}
  \label{eq:bending}
  (\ve x_{i+1}-\ve x_{i})(\ve x_{i}-\ve x_{i-1}) \ge (2a)^2 \cos \beta_{max}\;.
\end{equation}
Naturally, beads cannot overlap with the surface ($z_i>a$) or with each other
$\left| \ve x_i - \ve x_j \right| >2a$.

We describe the hydrodynamics using the mobility matrix formalism. If the
force acting on bead $i$ is $\ve F_i$, the resulting velocities are 
\begin{equation}
  \label{eq:mobility}
  \frac d{dt} \ve x_i= \sum_j M_{i,j} \ve F_j\;.
\end{equation}
In this formalism, each element $M_{i,j}$ is itself a $3\times 3$ matrix,
corresponding to 3 spatial dimensions.  In general, the above equation should
also include angular velocities and torques, but they are negligible for small
beads when the surface speeds due to rotational motion are much smaller than
those due to translational motion.  The mobility matrix is symmetric and
positive-definite \cite{Happel.Brenner}. Therefore, one can always invert it
to determine the friction matrix $\Gamma=M^{-1}$, which determines the forces
on particles moving with known velocities
\begin{equation}
  \label{eq:friction}
  \ve F_i= \sum_j \Gamma_{i,j} {\dot{\ve x}}_j \;.
\end{equation}

If the particles were at large distances relative to their sizes, the elements
of the mobility matrix would be determined by Blake's tensor
\cite{Blake.1971}, which represents the Green function of the Stokes flow in
the presence of a no-slip boundary. In our case the condition of large
interparticle distances is not fulfilled and we use the next higher
approximation, which is the Rotne-Prager tensor in the presence of a boundary,
as described in a previous paper \cite{Vilfan.Babic2010}.

The volume flow rate in $x$ direction, averaged over one beat period $T$,
depends on $x$-components of forces acting on particles and their heights $z$
above the boundary \cite{Smith.Gaffney2008}:
\begin{equation}
  \label{eq:volflow}
  Q=\frac 1 T \int_0^T \frac 1 {\pi \eta} \sum_i z_i(t) F_{x,i}(t) dt\;.
\end{equation}
The dissipation rate is simply the total power needed to move the beads
against viscous drag,
\begin{equation}
  \label{eq:dissipation}
  P=\sum_i \dot{\ve x}_i \cdot {\ve F}_{i}\;.
\end{equation}

We numerically maximized the quantity $Q^2/P$ for a set of angles
$\beta_{max}$ and different numbers of beads. We used the sequential quadratic
programming algorithm (SQP) from NAG numerical libraries (Numerical Algorithms
Group).  The full details of the numerical procedure are given in Appendix 3.

To study the collective efficiency and metachronal coordination, we studied an
array of $N_a \times N_a$ cilia (unit cell) on a square lattice with a lattice
constant $d$. We introduced periodic boundary conditions by adding
hydrodynamic interactions between particles and the representations beyond
lattice boundaries. So if a certain element in the mobility matrix describing
interaction between particles at $\ve x_i$ and $\ve x_j$ is $M_{i,j}(\ve x_i,
\ve x_j)$, we replace it by $M'_{i,j}(\ve x_i,\ve
x_j)=\sum_{p,q=-\infty}^\infty M_{i,j}(\ve x_i, \ve x_j+p A \hat e_x + q A
\hat e_y)$. Here $A=N_a d$ denotes the size of the unit cell. For the sake of
numerical efficiency, we used the full Rotne-Prager form for the first $o$
instances ($p,q=-o,\ldots o$) and approximated the interaction with its long
range limit, independent of the actual particle positions, for the rest (SI).

We expect the optimal solution to have the form of metachronal waves with a
wave vector $\ve k= (k_x,k_y)= (2\pi/A) (\kappa_x,\kappa_y)$. In order to
satisfy the periodic boundary conditions, $\kappa_x$ and $\kappa_y$ have to be
integer numbers, e.g., between $0$ and $N_a-1$.

\begin{figure*}
\begin{center}
\includegraphics{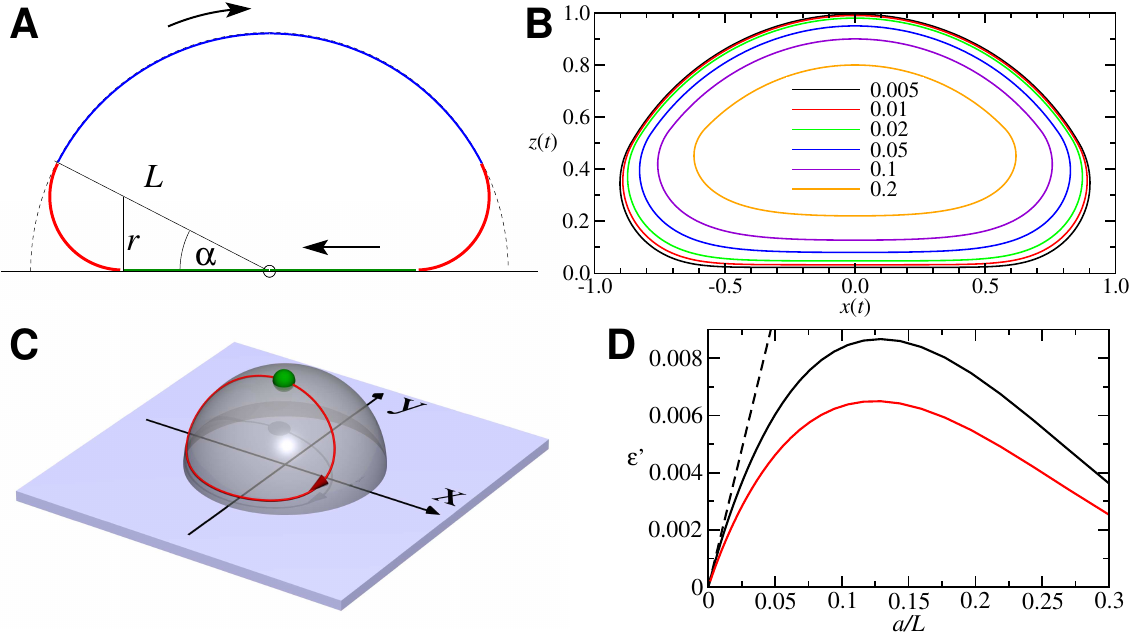}
\end{center}
\caption{Optimal trajectories of the 1-particle model. A) Idealized case of a
  small particle restricted to $\left| x \right| < L$. The solution consists
  of piecewise circular arcs, determined by geometric parameters $\alpha$ and
  $r/L$. B) Numerical solutions for finite-sized particles, plotted for
  different ratios $a/L$.  C) Optimal path for a particle at a constant
  distance from the origin, $\left| x \right|=L-a$, with $a=0.1\,L$. The
  transparent hemisphere symbolizes the surface on which the particle can
  move. D) Dimensionless efficiency $\epsilon'$ as a function of the
  dimensionless particle radius $a/L$. The black line shows the model with
  variable distance and the red line with a fixed distance from the
  origin. The dashed line shows the limit of small radii ($a\ll L$),
  $\epsilon'=0.192 \, a /L$.}
\label{2dmodel}
\end{figure*}

\section{Results}
\subsection{Single-particle model}

We first start with some simple models that are not necessarily feasible in
practice, but allow important insight into how the optimum is achieved. We
will follow the spirit of the model used to study the synchronization of cilia
\cite{vilfan2006a}, where we replace the cilium by a small spherical
particle. There are many swimmer models building on similar assumptions, for
example the three-sphere-swimmer \cite{Najafi.Golestanian.2004}, and they all
have in common that they assume the connections between spheres to be very
thin and neglect any hydrodynamic forces acting on them.

So the first hypothetical model we study is a single sphere of radius $a$ that
can move along an arbitrary path $\ve x(\omega t)$ in the half space above the
boundary, but in order to mimic the tip of a cilium it has to stay within the
distance $L$ of the origin, $\left| \ve x \right| \le L-a$. In order to
simplify the calculation we also assume that the sphere is small, $a\ll L$.
In this limit, we can neglect the effect of the boundary on the hydrodynamic
drag, which is then always $\gamma=6\pi \eta a$. The dissipated power is then
simply $P=\gamma {\dot {\ve x}}^2$.  Because it has to be constant in time, we
can also write it as
\begin{equation}
  \label{eq:hyp2}
  P=\gamma \ell^2/T^2
\end{equation}
with $\ell$ denoting the total distance traveled within one cycle and $T$ its
period.  The average volume flow follows from Eq.~(\ref{eq:volflow}) as
\begin{equation}
  \label{eq:hyp3}
  Q=\frac 1 {\pi \eta T} \int_0^T z(t) \gamma \dot x(t) dt=\frac {6a} T \oint z\,
  dx=\frac {6 S a}{T}
\end{equation}
where $S$ is the area of the particle trajectory, projected onto the $x-z$
plane.  The resulting efficiency is (\ref{eq:eff})
\begin{equation}
  \label{eq:hyp4}
  \epsilon=\frac {Q^2}{P}=\frac{6 S^2 a}{\pi \eta \ell^2}\;.
\end{equation}
To find the optimal path, we thus have to maximize the area-to-circumference
ratio of the path, while fulfilling the constraints $z>0$ and $\left| \ve x
\right| \le L$.  Obviously, there is no benefit in going out of the $x-z$
plane, but there is cost associated with it. Therefore, the optimum
trajectories will be planar. As any curve that minimizes its circumference at
a fixed surface area, the unconstrained segments of the trajectory have to be
circle arcs. The curve has the shape shown in Fig.~\ref{2dmodel}A. A numerical
solution shows that the area-to-circumference ratio is maximal if the angle
$\alpha$ defined in Fig.~\ref{2dmodel}A has the value $\alpha=0.483$.  The
resulting maximal efficiency in the limit $a\ll L$ is $\epsilon=0.192 \, L^2 a
/ \eta$, or, in dimensionless form, $\epsilon'=0.192 \, a /L$.

Solutions for finite values of $a/L$ are shown in Fig.~\ref{2dmodel}B and
their efficiencies in Fig.~\ref{2dmodel}D. The highest possible numerical
efficiency of this model is $\epsilon'= 0.0087$, which is achieved at
$a/L=0.13$.

Another version of the single-particle model is one in which the particle has
to maintain a fixed distance ($L-a$) from the origin, while it is free to move
along the surface of a sphere (Fig.~\ref{2dmodel}C).  This is an additional
constraint and can therefore only reduce the achievable efficiency.  As shown
by the red line in Fig.~\ref{2dmodel}D, the efficiency indeed lies somewhat
below that of the model with a variable distance and reaches a maximum value
of $\epsilon'= 0.0065$.

\subsection{N particles, stiff cilium}

The next minimalistic model we will study is a stiff cilium: a straight chain
of $N$ beads with radius $a$ and a total length of $L=2Na$ that can rotate
freely around the center of the first bead. The problem is related to
artificial cilia driven by a magnetic field
\cite{Vilfan.Babic2010,Gauger.Stark2009,Downton.Stark2009} in which the
orientation of the cilium largely (although not completely) follows the
direction of the magnetic field. A related optimization has been performed by
Smith et al.\ \cite{Smith.Gaffney2008}, but with two important
differences. First, Smith et al.\ optimize the volume flow alone and not the
efficiency. Their optimal stroke therefore touches the surface during the
recovery stroke, while ours has to keep some distance in order to limit the
dissipation.  Second, they restrict themselves to cilia beating along tilted
cones, whereas we allow any arbitrary pattern.

\begin{figure}
\begin{center}
\includemovie[
        poster=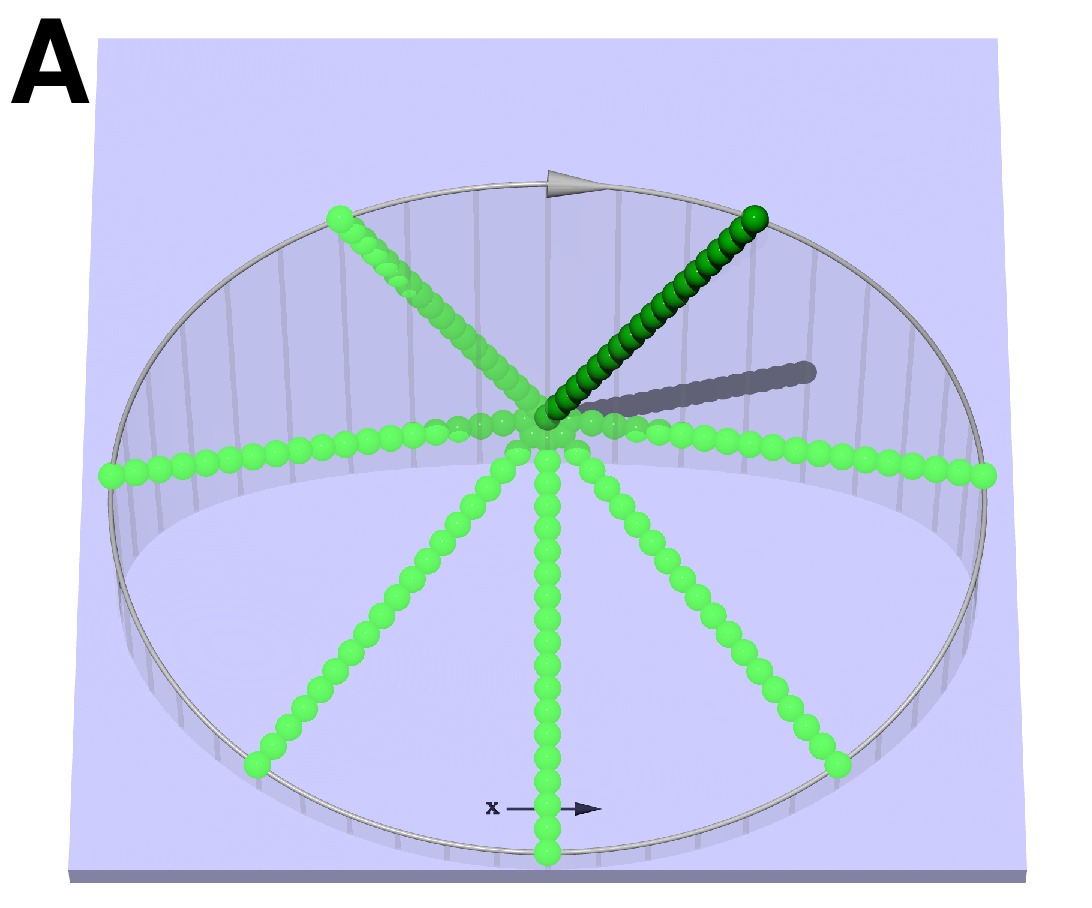,
	label=figure2a.u3d,
	3Daac=50.000000, 3Droll=0.000000, 3Dc2c=0 -3 5,
        3Droo=300.092304, 3Dcoo=0.0 0.0 0.0,
	3Dlights=CAD,
]{6cm}{5cm}{Fig2A.u3d}
\\
\includemovie[
        poster=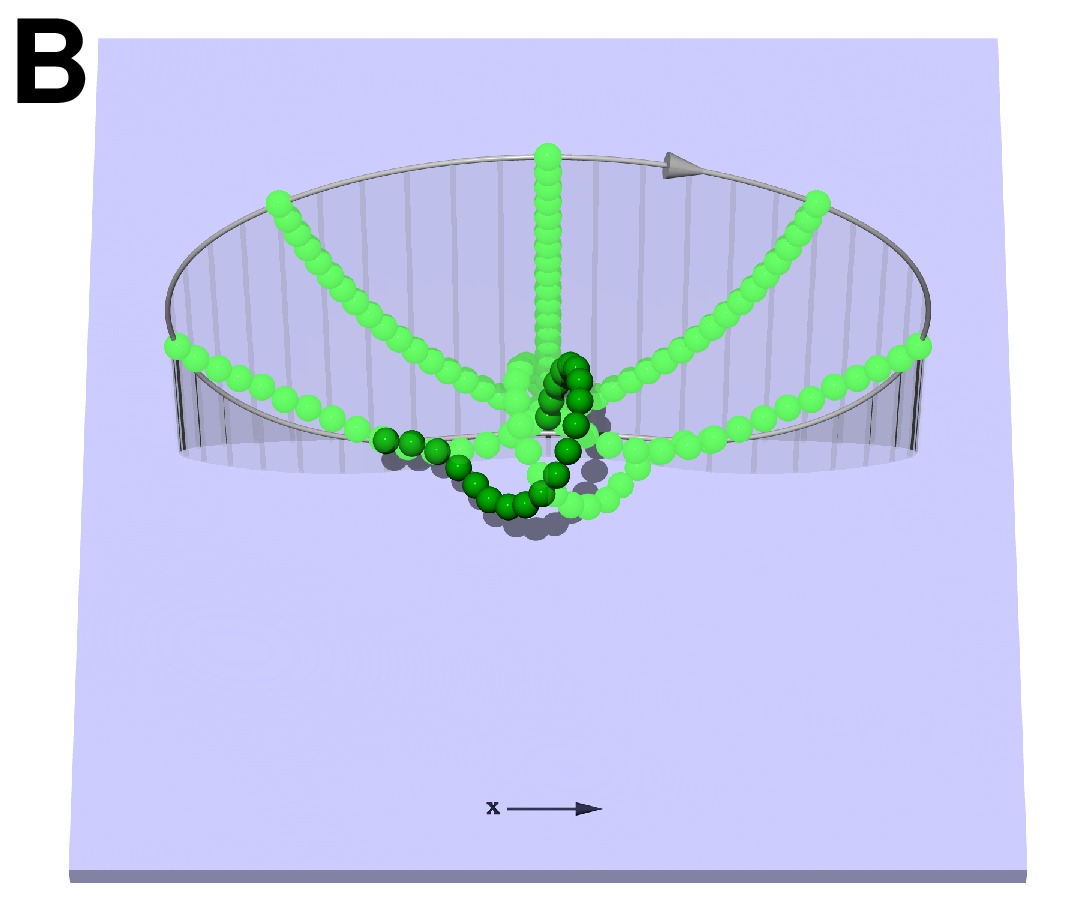,
	label=figure2b.u3d,
	3Daac=50.000000, 3Droll=0.000000, 3Dc2c=0 -3 5,
        3Droo=300.092304, 3Dcoo=0.0 0.0 0.0,
	3Dlights=CAD,
]{6cm}{5cm}{Fig2B.u3d}
\\
\includemovie[
        poster=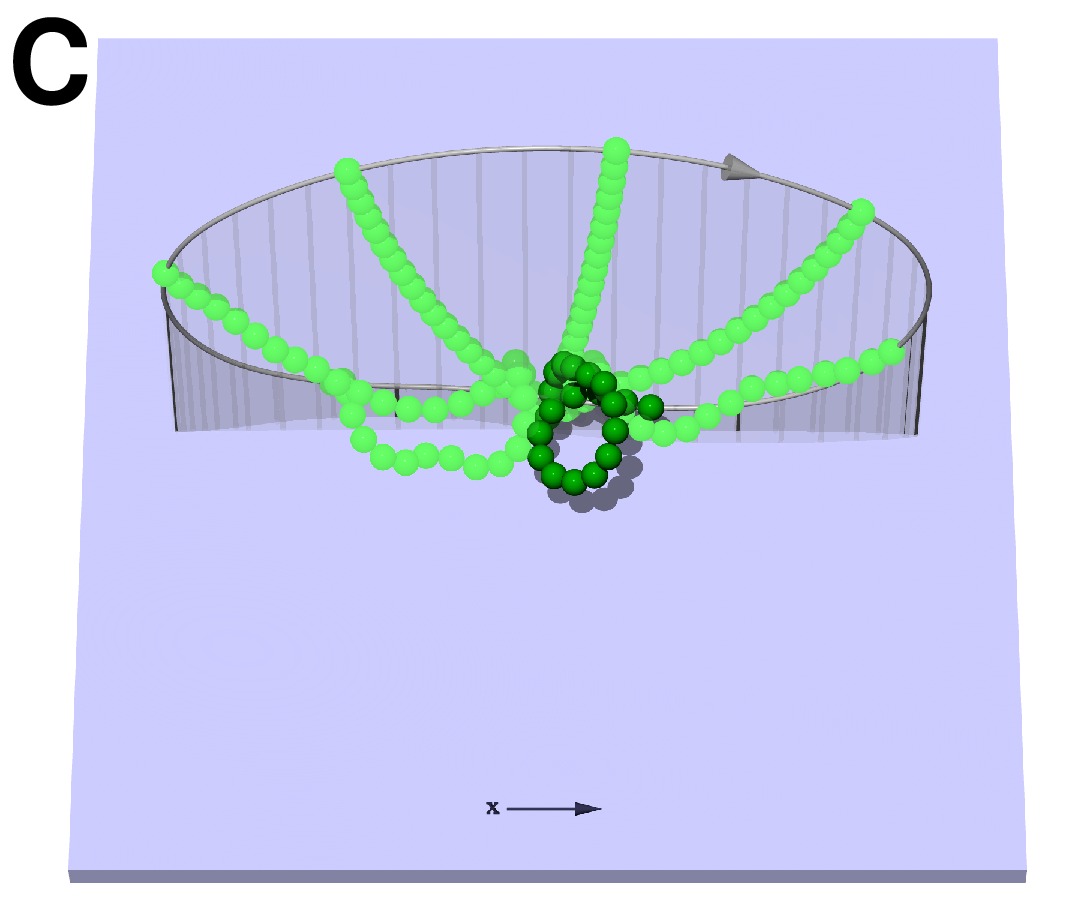,
	label=figure2c.u3d,
	3Daac=50.000000, 3Droll=0.000000, 3Dc2c=0 -3 5,
        3Droo=300.092304, 3Dcoo=0.0 0.0 0.0,
	3Dlights=CAD,
]{6cm}{5cm}{Fig2C.u3d}
\\
\includegraphics[width=5.5cm]{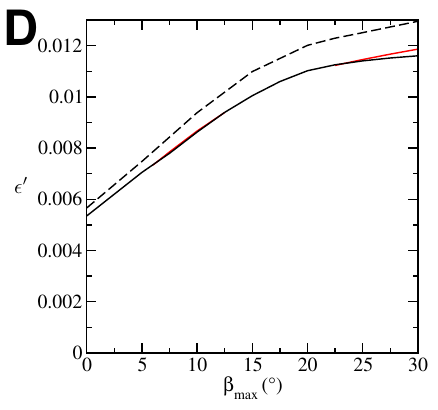}
\end{center}
\caption{Optimal beating patterns of a cilium consisting of $N=20$ particles
  with different allowed bending angles $\beta_{\rm max}$.  The gray surface
  shows the projection of the tip trajectory on the $x-y$ plane. A) A stiff
  cilium, $\beta_{\rm max}=0$. B) A flexible cilium, $\beta_{\rm
    max}=20^\circ$. The optimal stroke is symmetric in $x$ direction. C)
  Flexible cilium, $\beta_{\rm max}=30^\circ$. The symmetry in $x$ direction
  is broken. D) Dimensionless efficiency $\epsilon'$ as a function of
  $\beta_{\rm max}$. The black line shows the optimal symmetric solution and
  the red line the asymmetric solution in cases where it is more efficient.
  The dashed line shows the maximum efficiency for $N=10$ and double
  $\beta_{\rm max}$ (corresponding to the same curvature).}
  \label{fig:examples}
\end{figure}

The motion of a stiff cilium on its optimal trajectory is shown in Figure
\ref{fig:examples}A. The path of its tip closely resembles that of a single
sphere at a fixed radius.  The resulting dimensionless efficiency for $N=20$
beads is $\epsilon'\approx 0.00535$.

\subsection{N particles, flexible cilium}

As the next level of complexity, and at the same time the first realistic
description of biological cilia, we now study a flexible cilium consisting of
$N$ spherical particles (we use $N=10$ and $N=20$). The bending angle per
particle is restricted to $\beta_{\rm max}$ (\ref{eq:bending}).  Such a
constraint is necessary for two reasons. For one, the curvature of a
biological cilium is restricted by the bending rigidity of the axoneme. In
addition, our $N$-particle model veritably represents a continuous cilium only
if the curvature radius is sufficiently larger than the size of a
sphere. Examples of beating patterns obtained by numerical optimization are
shown in Figure \ref{fig:examples}B,C. Figure \ref{fig:examples}D shows the
dimensionless efficiency $\epsilon'$ as a function of $\beta_{\rm max}$.

It is instructive to look at fundamental symmetries of the problem at this
point. First, as any of the problems studied here, it is symmetric upon
reflection $y\to -y$. For every clockwise beat, there is an equivalent
counterclockwise beat with the same efficiency. All cycles discussed here
spontaneously break the $y$ symmetry. In addition, the equations are invariant
upon reflection $x\to -x$ with simultaneous time reversal, $t\to -t$.  This
symmetry can be broken or not at the efficiency maximum.  Interestingly, this
depends on the allowed bending between adjacent elements $\beta_{\rm max}$.
For example, the solution shown in Fig.~\ref{fig:examples}B is $xt$-symmetric,
while the one in Fig.~\ref{fig:examples}C is not. 

\begin{figure*}
  \begin{center}
\includegraphics{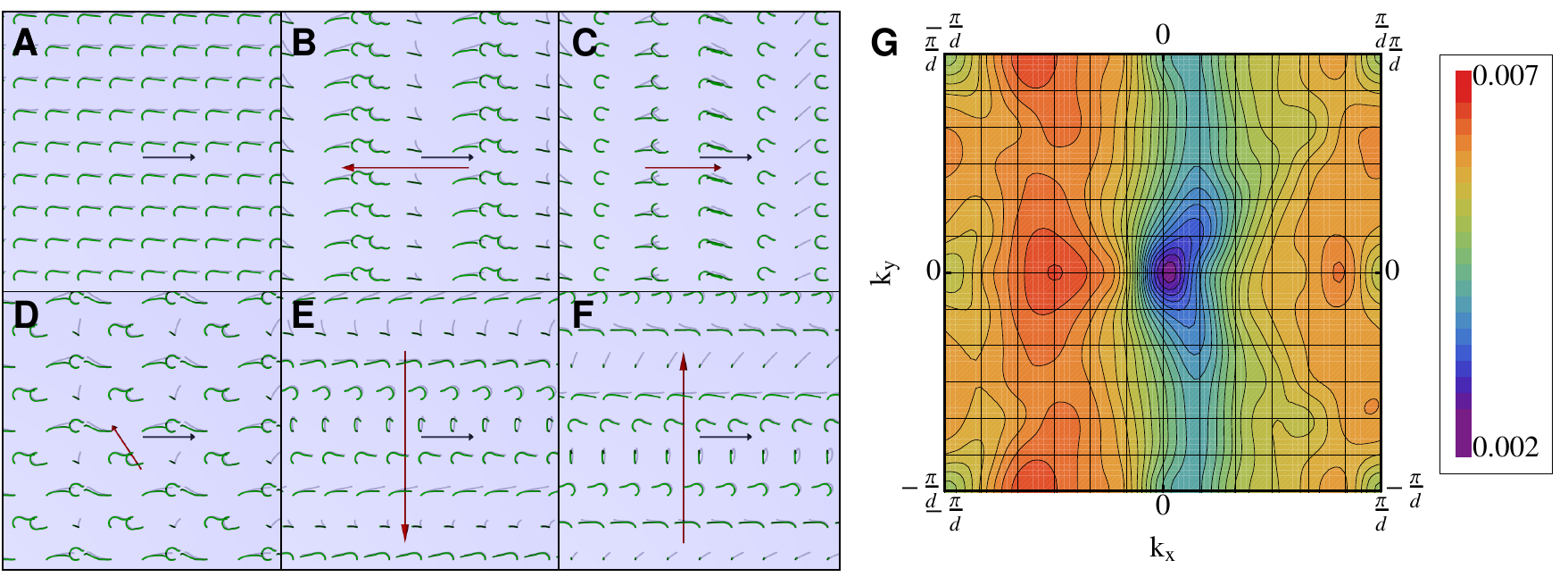}
  \end{center}
  \caption{Optimal solutions at fixed wave vectors for interciliary distance
    $d=1.0\times L$, $N=20$, $\beta_{\rm max}=15^\circ$ and $N_a=12$.  A-F)
    Optimal solutions for wave vectors $(k_x,k_y)= (0,0)$ (A), $(-\pi/(2d),0)$
    (B), $(5\pi/(6d),0)$ (C), $(-2\pi/(3d),\pi/d)$ (D), $(0,-\pi/(3d))$ (E)
    and $(0,\pi/(3d))$ (F). The blue arrow ($x$ axis) denotes the direction of
    pumping and the red arrow the wavelength and direction of metachronal wave
    propagation.  G) Efficiency $\epsilon_c'$ (red color represents high
    efficiency) as a function of the wave vector $(k_x,k_y)$. The maximum
    efficiency is in this case achieved for $\ve k=(-\pi/(2d),0)$ and
    antiplectic waves are generally more efficient than symplectic. The
    synchronous solution $(0,0)$ represents the global minimum of efficiency.}
  \label{fig:contourplot}
\end{figure*}

\subsection{Multiple cilia and metachronal waves}

\begin{figure*}
\begin{center}
\includegraphics{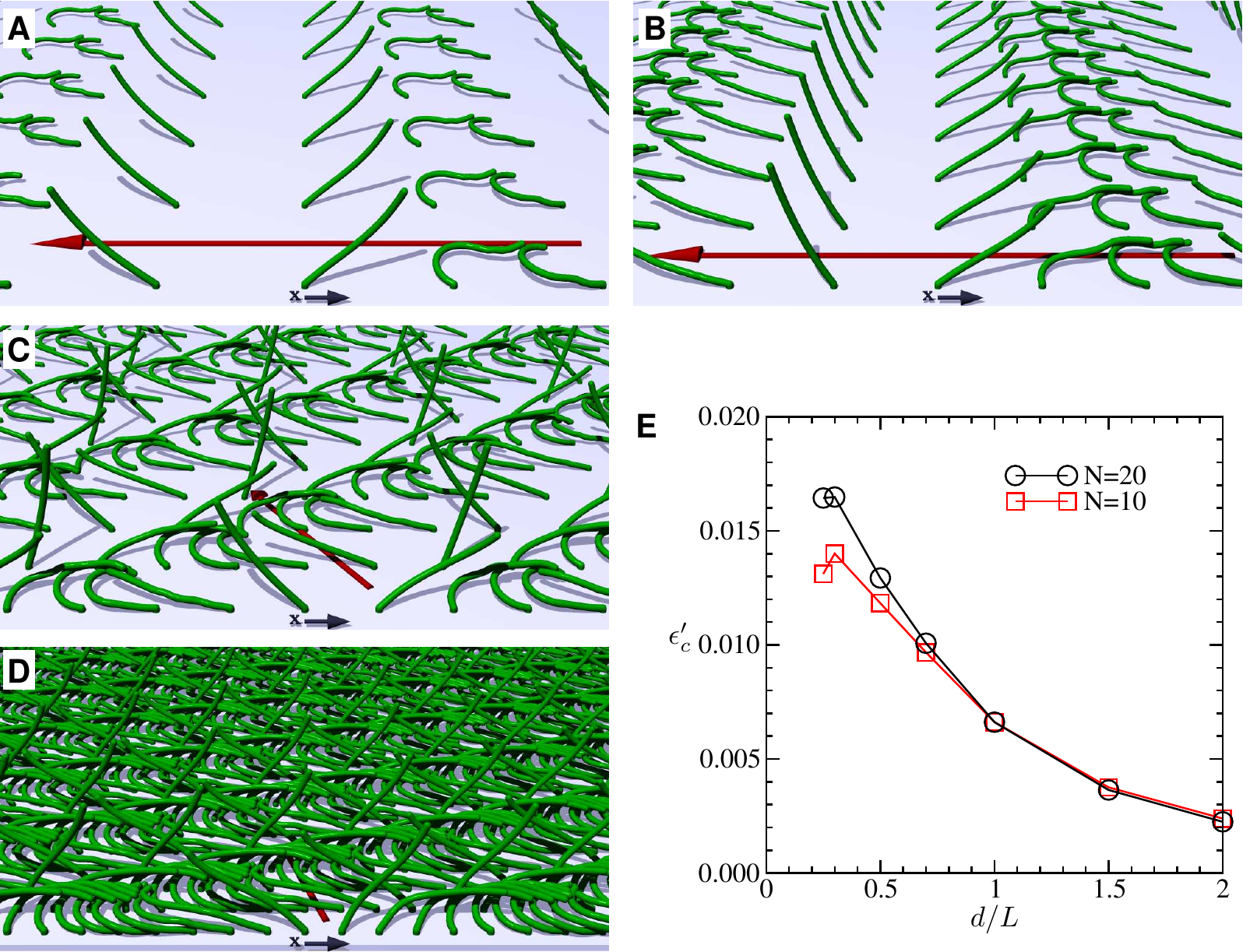}
\end{center}
\caption{Optimal solutions for various interciliary distances: $d=1.0\,L$ (A),
  $d=0.7\,L$ (B), $d=0.5\,L$ (C) and $d=0.25\,L$ (D). For reasons of clarity
  the front rows of cilia are omitted and instead of individual spheres used
  in the calculation a tube connecting them is shown. E) Highest efficiency
  $\epsilon_c'$ as a function of the interciliary distance $d$ for cilia
  consisting of $N=20$ (circles) and $N=10$ (squares) spheres. For $N=10$, we
  set $\beta_{\rm max}=30^\circ$ in order to allow the same maximum
  curvature.}
\label{fig:snapshots}
\end{figure*}

We solve the optimization problem of $N_a\times N_a$ cilia ($N_a=12$) with
periodic boundary conditions by imposing a wave vector $(k_x, k_y)$, finding
the optimal solution for that vector and repeating the procedure for $N_a
\times N_a$ wave vectors. Examples of optimal solutions for 6 different
wave vectors are shown in Fig.~\ref{fig:contourplot}A-F.  The efficiency
$\epsilon_c'$ as a function of the wave vector is shown in Fig.\
\ref{fig:contourplot}G.  Note that all solutions determined in this section
are for counterclockwise beating (viewed from above). For clockwise strokes
the $y$ component of the wave vector would have the opposite sign. Optimal
solutions for four different values of the interciliary distance $d$ are shown
in Figure \ref{fig:snapshots}A-D and the optimal efficiency as a function of
$d$ in Figure \ref{fig:snapshots}E.

Figure \ref{fig:contourplot}G shows that the efficiency depends more strongly
on the longitudinal ($k_x$) component of the wave vector than on the lateral
($k_y$). This could partly be due to the nature of the hydrodynamic
interaction, which is stronger in longitudinal direction, and partly because
the cilia exert larger motion in longitudinal direction and therefore come
closer to their neighbors along the $x$ axis. Antiplectic metachronal waves
(waves propagate in the opposite direction from the fluid pumping) generally
have a higher efficiency than symplectic, but the fine structure is much more
complex. For low ciliary densities, the optimal solution is found for waves
propagating along the $x$ axis. However, for higher densities solutions with a
positive $k_x$ are more efficient, which means that the waves are
dexio-antiplectic. Efficiency also grows with increasing density. But when the
interciliary distance reaches $d=0.25\,L$ crowding becomes a problem and the
efficiency drops again. At even higher densities the solution becomes
increasingly difficult to find because of the complicated topology of densely
packed cilia. Another problem is that the wavelength of the optimal solution,
relative to the lattice constant, becomes increasingly long at high densities,
which would require a unit cell larger than $12\times12$ used in our
calculations.

\subsection{Swimming efficiency of a ciliated microorganism}

We can finally use these results to estimate the maximum possible swimming
efficiency of a ciliated microorganism. For the sake of simplicity, we assume
that the swimmer has a spherical shape with radius $R$. According to
Lighthill's definition, the swimming efficiency is defined as
\begin{equation}
  \label{eq:lighthill}
  \epsilon_L=\frac{6\pi \eta R V^2}{P_{\rm tot}}
\end{equation}
where $V$ is the velocity and $P_{\rm tot}$ the total dissipated power
\cite{Lighthill1952}. Assuming that the layer of cilia is thin in comparison
with the size of the organism ($L\ll R$), the swimming velocity $V$ can be
calculated as \cite{Stone.Samuel1996,Julicher.Prost2009}
\begin{equation}
  \label{eq:swimmingvelocity}
  V=\frac 1{4\pi R^2}\int {\mathbf v} d^2s=\frac 1 {4\pi} \int_0^\pi
  v(\theta) 2\pi \sin^2 \theta d\theta \;.
\end{equation}
Here $v(\theta)$ is the propulsion velocity above the ciliated layer.  The
dissipation can be expressed as
\begin{equation}
  \label{eq:totdissipation}
  \begin{split}
  P_{\rm tot}=\int \rho P d^2s =  \int_0^\pi
  \rho P(\theta) 2\pi R^2 \sin \theta d\theta \\= \frac {2 R^2 \eta}{\pi \epsilon_c' L} \int_0^\pi
  v^2(\theta) \sin \theta d\theta\;.
\end{split}
\end{equation}
In the second equality we used the definition (\ref{eq:eff-coll}), as well as
the relationship between $v$ and $\rho Q$ (\ref{eq:vinfty}).  In order to
obtain the maximum $V$ at a given $P_{\rm tot}$, the two functional
derivatives should be related through a Lagrange multiplier
\begin{equation}
  \label{eq:lagrange}
  \frac {\delta V[v]}{\delta v(\theta)} = \lambda  \frac {\delta P_{\rm tot}[v]}{\delta v(\theta)} \;,
\end{equation}
which is fulfilled if the angular dependence has the form $v(\theta)=v_0
\sin\theta$. This leads to
\begin{equation}
  \label{eq:swimmingvelocity1}
  V=\frac 2 3 v_0
\end{equation}
and 
\begin{equation}
  \label{eq:dissipation1}
  P_{\rm tot}=\frac{8\eta R^2 v_0^2}{3\pi \epsilon_c' L}\;.
\end{equation}
Together, these equations give Lighthill's efficiency 
\begin{equation}
  \label{eq:lighthill1}
  \epsilon_L=\pi^2 \frac L R \epsilon_c' \;.
\end{equation}
With a maximum $\epsilon_c'\approx 0.016$ and a typical ratio $L/R\approx 0.1$,
we obtain $\epsilon_L\approx 0.016$. 

For comparison, an optimized envelope model yields an efficiency around
$3\,\%$ with the same parameters (if we translate the ciliary length into
maximal displacement of the envelope) \cite{Michelin.Lauga2010}, which shows
that the latter is a relatively good approximation for cilia if they operate
close to the optimal way.

\section{Discussion}

We introduced a scale invariant efficiency measure for the fluid pumping by
cilia and started with optimizing three simple instructive systems: a free
sphere, a sphere at a constant distance from the origin and a stiff cilium. Of
those three, the free sphere can reach the highest efficiency. But they are
all topped by a flexible cilium. The thickness of a cilium has only a small
effect on its efficiency, which strengthens our choice to describe the cilium
as a chain of beads. Depending on the allowed bending, the flexible cilium can
have different shapes of the optimal beating pattern. In most cases the cilium
curls up during the recovery stroke, rather than sweeping along the surface.
Such shapes appear ``unnatural'' if we compare them with those observed in
microorganisms \cite{Brennen.Winet1977}.

But the collective optimization of ciliary carpets leads to beating patterns
that are strikingly similar to what is observed in many ciliated
microorganisms. Unlike isolated cilia, they contain a recovery stroke during
which they sweep along the surface.  This is primarily due to the fact that
beating patterns that are optimal for a single cilium (e.g., as shown in
Fig.~\ref{fig:examples}B) are not possible on a dense grid due to steric
hindrance. The sweeping recovery stroke, on the other hand, allows dense
stacking of cilia (best seen in Fig.~\ref{fig:snapshots}D) which further
reduces drag as well as backward flow.  The optimal effective stroke becomes
significantly faster than the recovery stroke. While a single cilium reaches
its highest efficiency if the effective stroke takes about $45\%$ of the
cycle, the optimum is around $20-25\%$ for densely packed cilia. A similar
ratio has been observed in \textit{Paramecium} \cite{Gueron.Blum1997}.  The
distance between adjacent cilia in \textit{Paramecium} is between $0.15\,L$
and $0.25\,L$ \cite{Sleigh1969}, consistent with the predicted optimum around
$d=0.25\,L$.
It is interesting that the efficiency of any other wave vector is higher than
the efficiency of the synchronous solution (wave vector 0). This is in
agreement with some previous simpler, one dimensional models
\cite{Gauger.Stark2009}, but has not yet been shown on a 2-D lattice. We also
find that antiplectic waves are generally more efficient than symplectic,
although symplectic solutions with a relatively high efficiency exist, too.
For high densities and cilia beating counterclockwise, the waves become almost
dexioplectic (meaning that the effective stroke points to the right of the
wave propagation) and the wavelength becomes similar to the cilium length $L$
- both findings are in agreement with observations on \textit{Paramecium}
\cite{Machemer1972,Sleigh74}. For cilia beating clockwise, laeoplectic waves
would be more efficient, which is indeed observed
\cite{Gheber.Priel1990a}. Although the effect of thickness is small, it is
interesting to note that thicker cilia have a slightly higher efficiency when
isolated or at low surface densities, but are outperformed by thinner cilia at
high densities.

The total energetic efficiency of swimming in \textit{Paramecium} has been
measured as $0.078\,\%$ \cite{Katsu-Kimura.Mogami2009}. This figure includes
losses in metabolism and force generation -- the hydrodynamic swimming
efficiency alone has been estimated as $0.77\,\%$. This comes close (by a
factor of 2) to our result for the maximally possible Lighthill efficiency of
a spherical ciliated swimmer $\epsilon_L\approx 0.016$.  A biflagellate
swimmer like \textit{Chlamydomonas} has a lower theoretical efficiency of
$0.008$ \cite{Tam.Hosoi2011}, but it is still within the same order of
magnitude.

Although efficiencies below $1\,\%$ seem low, we have shown that
\textit{Paramecium} still works remarkably close to the maximum efficiency
that can be achieved with its length of cilia. While longer cilia might have a
higher swimming efficiency, there are other considerations that are not
included in this purely hydrodynamic study. For example, the bending moments
and the power output per ciliary length can be limiting \cite{Sleigh.Blake1977}.
Thus, our study shows that at least for ciliates like \textit{Paramecium},
Purcell's view that efficiency is irrelevant for ciliary propulsion has to be
revisited.  Efficiency of swimming does matter for them, and in their own
world they have well evolved to swim remarkably close to the optimal way.

\acknowledgments
  We have benefited from fruitful discussions with Frank Jülicher.  This work
  was supported by the Slovenian Office of Science (Grants P1-0099, P1-0192,
  J1-2209, J1-2200 and J1-0908).

\appendix
\setcounter{equation}{0}
\setcounter{figure}{0}
\setcounter{table}{0}
\renewcommand{\theequation}{S\arabic{equation}}
\renewcommand{\thetable}{S\arabic{table}}
\renewcommand{\thefigure}{S\arabic{figure}}
\section{Appendix 1. Justification of the efficiency criterion}

In this section we prove that instead of minimizing the dissipated power $P$
for a constant volume flow rate $Q$ we can maximize the numerical efficiency
$\epsilon=Q^2/P$ at a constant beating frequency $\omega$.  Let the volume
flow rate $Q[\ve x(\omega t)]$ and dissipation $P[\ve x (\omega t)]$ be
functionals of the trajectory shape $\ve x(\omega t)$ and functions of
$\omega$. The efficiency $\epsilon$, defined as $\epsilon[\ve x (\omega
t)]=Q^2/P$, is only a functional of $\ve x(\omega t)$, but independent of
$\omega$. The relationship
\begin{equation}
  \label{eq:functional}
-\frac {Q^2}{P^2} \left.\frac{  \delta P }{\delta \ve x}\right|_Q =\left.\frac{  \delta \epsilon }{\delta \ve x}\right|_Q= \left.\frac{  \delta
    \epsilon }{\delta \ve x}\right|_\omega + \left.\frac{  \partial \epsilon
  }{\partial \omega }\right|_{\ve x} \left.\frac{  \delta
    \omega }{\delta \ve x}\right|_Q= \left.\frac{  \delta
    \epsilon }{\delta \ve x}\right|_\omega
\end{equation}
proves immediately that minimizing the dissipated power while keeping the
volume flow $Q$ constant is equivalent to maximizing $\epsilon$ at a constant
$\omega$.  

\section{Appendix 2. Optimal solutions have a constant dissipation
}
In the following we demonstrate that a trajectory with optimal efficiency
always has a dissipation that is constant in time. The pumping performance
only depends on the shape of the trajectory and the period $T$, but not on the
velocity along the trajectory. We have to determine the latter in a way that
minimizes the average dissipation. We parameterize the trajectory as $\ve
x(\varphi)$ where the phase $\varphi$ is a function of time that fulfills
$\varphi(0)=0$ and $\varphi(T)=2\pi$.  The dissipation at any moment is
quadratic in velocity, $P(t)=\dot {\ve x}^T \gamma(\ve x) \dot {\ve x}$, and
can be written as
\begin{equation}
P(t)=(d \ve
x/d\varphi)^T \gamma(\ve x)(d \ve x/d\varphi) \dot \varphi ^2 =p(\varphi)\dot
\varphi ^2\;.
\end{equation}
We obtain the average dissipation by integrating over one period,
\begin{equation}
\begin{split}
  \bar P&=\frac 1 T \int_0^T P(t) dt= \frac 1 T \int_0^T p(\varphi)
  \dot\varphi^2 dt\\&= \frac 1 {T} \int_0^{2\pi} p(\varphi)
  \left( \frac{dt}{d\varphi} \right) ^{-1} d\varphi= \frac 1 {T}
  \int_0^{2\pi} \tilde p(\varphi, t') d\varphi\;,
\end{split}
\end{equation}
with $t'=dt/d\varphi$ and $ \tilde p(\varphi, t')=
p(\varphi)/t'$. Minimization of the average dissipation while keeping the
period constant leads to the following Euler-Lagrange equation (note the
swapped roles of $t$ and $\varphi$)
\begin{equation}
  \label{eq:el} 0=\frac{\partial \tilde p}{\partial t} -\frac{d}{d\varphi}
\frac{ \partial\tilde p}{\partial t'}=\frac{d}{d\varphi} \frac {p}{
{t'}^2}=\frac{dP}{d\varphi}\;.
\end{equation} This proves that the optimal solution has a dissipation $P$
that is constant in time.

\section{Appendix 3. Numerical optimization procedure}

\subsection{Single cilium}

For a single cilium, modeled as a chain of $N$ beads, we parameterize the
stroke as a sequence of $N_S$ equally spaced time steps with duration $\Delta
t=T/N_S$. Let the vector ${\bf x}_i(\tau)$ represent the position of the bead
$i$ at time step $\tau$.  Because the trajectory is periodic in time, we have
${\bf x}_i(\tau)= {\bf x}_i(\tau+N_S)$.  For a given stroke the volume flow
rate is calculated according to Eq. \ref{eq:volflow} as
\begin{equation} 
\begin{split}
Q=\frac 1 {\pi \eta N_S} \sum_{\tau=1}^{N_S}\sum_{i,j=1}^{N}
\frac{z_{i}( \tau+1)\Gamma_{i,j}(\tau+1)+z_{i}( \tau) \Gamma_{i,j}(\tau)}{2}
\,\\\times  \frac{{\bf x}_j(\tau+1)-{\bf x}_j(\tau)}{\Delta t}\;.
\end{split}
\end{equation} 
$\eta$ denotes the viscosity of the surrounding fluid and $z_i$ the height of
$i$-th bead above the surface.  The dissipated power, defined in
Eq. \ref{eq:dissipation}, can be written as
\begin{equation} 
\begin{split}
P=&\frac 1 {N_S}\sum_{\tau=1}^{N_S}\sum_{i,j=1}^{N}
\frac{\Gamma_{i,j}(\tau+1)+ \Gamma_{i,j}(\tau)}{2} \,\\ &\times \frac{\big({\bf
x}_i(\tau+1)-{\bf x}_i(\tau) \big) \big ({\bf x}_j(\tau+1)-{\bf x}_j(\tau)
\big)}{(\Delta t)^2} \;.
\end{split}
\end{equation} 
The dimensionless efficiency, which we want to maximize, follows as $\epsilon'
=L^{-3} \eta Q^2/P$ with $L=2aN$ (\ref{eq:eff-dl}).  The friction matrix
$\Gamma_{i,j}(t)$ that appears in above expressions is obtained by numerical
inversion of the mobility matrix $M_{i,j}$, which is calculated using the
Rotne-Prager approximation in the presence of a boundary, as described in
\cite{Vilfan.Babic2010}.

We computed the optimal strokes using the NAG (The Numerical Algorithms Group
Ltd., Oxford, UK) Fortran Library E04UGF routine, which maximizes an arbitrary
smooth function subject to constraints using a sequential quadratic
programming method.  At each time step we parameterize the bead positions $\ve
x_i$ by a set of $N-1$ pairs of polar ($\theta$) and azimuthal ($\phi$)
angles, as described in the main text (Section ``The Model'').  Therefore, the
parameter space in which we search for the for the efficiency maximum is $2
N_S (N-1)$ dimensional. For most calculations presented here ($N=20$,
$N_S=84$), this means 3192 dimensions. We still consider optimization in such
high-dimensional space preferable to parameterizing the beat shape in terms of
Fourier modes (see, e.g. \cite{Tam.Hosoi2011}), which would have less
variables, but therefore a more complex landscape with more local maxima.

The stroke shapes are subject to two types of constraints: (i) beads are
modeled as hard core objects that allow no overlapping of two beads or a bead
with the surface, (ii) the maximum bending angle defined between three
adjacent beads is limited to $\beta_{\rm max}$ (\ref{eq:bending}).

We initialized the optimization procedure with a simple beating pattern (a
tilted cone) that had the correct handedness (clockwise) and checked that the
result was otherwise independent of the initial state.

\subsection{Carpets of cilia}

For an array of cilia the size of parameter space would grow with the square
of the system size $N_a$ and the demand to calculate the hydrodynamic mobility
matrix with its fourth power.  The problem would become numerically unsolvable
even for a relatively small field of cilia. However, in an infinite system (or
a system with periodic boundary conditions) we expect the optimal solution to
have the form of metachronal waves with an unknown wave vector. The
optimization problem at a fixed wave vector then requires the same number of
parameters as the optimization of a single ciliary beat. The globally optimal
solution can be found by repeating the calculation for all wave vectors that
fulfill the periodic boundary conditions.

For a wave vector $(k_x,k_y)$, the $i$-th sphere of the cilium at lattice
position $(\alpha,\beta)$ follows the path
\begin{equation}
\ve x_{i,\alpha,\beta}(\omega t)= \alpha d\, \hat e_x +  \beta d\, \hat e_y +
\ve x_{i}(\omega t - \alpha d\, k_x -\beta d\, k_y)\;,
\end{equation}
where $\ve x_{i}(\omega t)$ is the path of the cilium at lattice position
$(0,0)$, which we are optimizing.  Similarly, the force acting on the same
bead has the following time dependence
\begin{equation}
\ve F_{i,\alpha,\beta}(\omega t)= \ve F_{i}(\omega t - \alpha d\, k_x -\beta d\, k_y)\;.
\end{equation}

The equations of motion of $i$-th bead belonging to the cilium $(0,0)$ read
\begin{equation}
\begin{split}
  \label{eq:mobilityarray}
  \frac d{dt}& \ve x_{i,0,0}(\omega t) = \frac d{dt} \ve x_{i} (\omega t)\\& =
  \sum_{j=1}^{N}\sum_{\alpha=-\infty}^\infty \sum_{\beta=-\infty}^\infty
  M_{i,j;\alpha, \beta} (\omega t) \ve F_j(\omega t- \alpha d\, k_x -\beta d\,
  k_y) \;.
\end{split}
\end{equation}
In this expression $M_{i,j;\alpha, \beta}(\omega t)$ represents the element of
the mobility matrix that links the response of $i$-th bead of the cilium
$(0,0)$ to the force acting on $j$-th bead of the cilium $(\alpha,\beta)$ at
time $t$.

\begin{figure}
\begin{center}
\includegraphics[width=8cm]{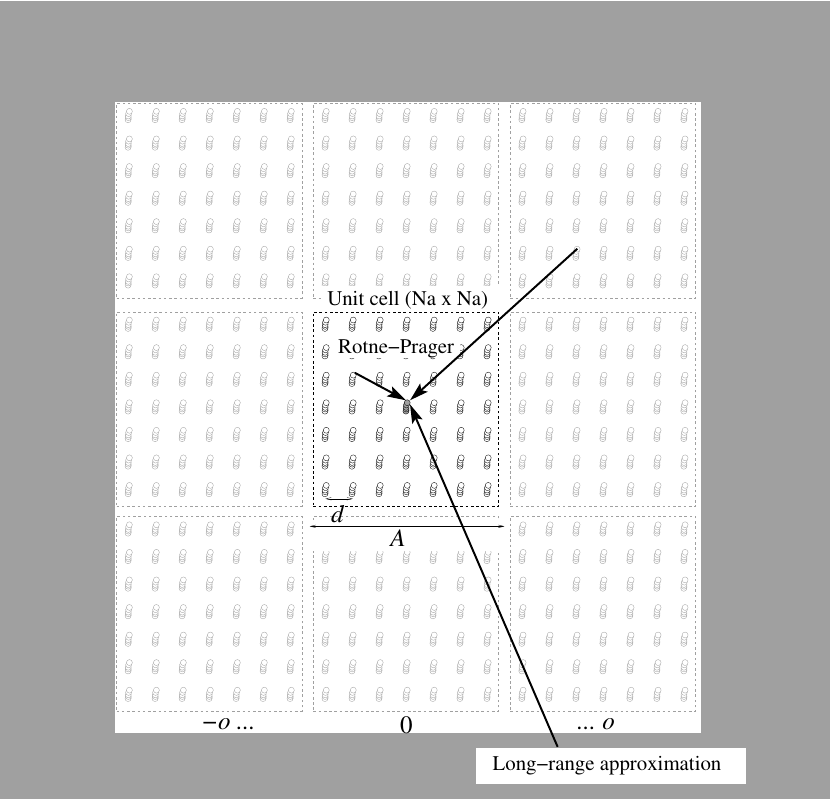}
\end{center}
\caption{The elements of the mobility matrix describe the response of the
  particles forming the central cilium ($0,0$) to forces acting on all other
  particles. Cilia are positioned on a square lattice with lattice constant
  $d$, while the unit cell comprising $N_a\times N_a$ cilia has the edge
  length $A$. To facilitate the calculation we use the full Rotne-Prager form
  for $(2o+1)\times (2o+1)$ unit cells and use the long-range limit
  expressions for more distant cilia.}
\label{fig:lattice}
\end{figure}

Periodic boundary conditions are fulfilled if $k_x$ and $k_y$ are multiples of
$2\pi/A$, with $A=N_a d$ denoting the size of the field
(Fig.~\ref{fig:lattice}). We can therefore use integer wave vectors
$\kappa_{x,y}=k_{x,y}A/(2\pi)$ instead.  We can make use of the periodicity by
carrying out the summation in the above equation
\begin{equation}
  \label{eq:mobilityarray1}
  \frac d{dt} \ve x_{i} (\omega t)= \sum_{j=1}^{N}\sum_{\alpha,\beta}
  M'_{i,j;\alpha, \beta} (\omega t) \ve F_j(\omega t- \alpha d\, k_x -\beta d\, k_y)
\end{equation}
with 
\begin{equation}
  \label{eq:mprime}
M'_{i,j;\alpha, \beta} =\sum_{p,q=-\infty}^\infty
M_{i,j;\alpha+p N_a,\beta+q N_a}\;.
\end{equation}
The indices $\alpha$ and $\beta$ run over one unit cell, centered
around the origin. For odd $N_a$, this would be from $-(N_a-1)/2$ to
$(N_a-1)/2$. For an even $N_a$, they run from $-N_a/2$ to $N_a/2$, but the
boundary terms are given half weight each. This distribution is necessary in
order to preserve the symmetry of the mobility/friction matrix.

For the sake of numerical efficiency, we use the full Rotne-Prager form for
the first $o$ instances ($p,q=-o,\ldots o$) and approximate the interaction
with
\begin{equation}
  \label{eq:m-approx}
\begin{split}
  M_{i,j;\alpha, \beta}&\approx \frac 3 {2\pi \eta} \frac{z_iz_j}{((pA+\alpha
    d)^2+(qA+\beta d)^2)^{5/2}} \\  &\hspace*{-0.5cm}\times
  \left( \begin{array}{ccc}(pA+\alpha d)^2&(pA+\alpha d ) (qA+\beta d)
      &0\\(pA+\alpha d) (qA+\beta d)  & (qA+\beta d)^2 & 0\\ 0 & 0 & 0
    \end{array} \right)
\end{split}
\end{equation}
for larger indices. We used $o=1$ for the optimization and checked the final
efficiency with $o=2$ (the correction was not significant). The contribution
of distant cilia towards $M'$ can easily be calculated numerically to
quadratic order in $\alpha$ and $\beta$ and has the form
\begin{equation}
 \frac {3 z_i z_j} {2\pi \eta A^3} \left( \begin{array}{ccc}C_1+C_2 \frac{\alpha^2}{N_a^2}
     +C_3 \frac{\beta^2}{N_a^2}& C_4 \frac{\alpha \beta}{N_a^2} &0\\ C_4 \frac{\alpha \beta}{N_a^2}  &
    C_1+C_2 \frac{\beta^2}{N_a^2}
     +C_3 \frac{\alpha^2}{N_a^2} & 0\\ 0 & 0 & 0
    \end{array} \right)\;.
\end{equation}
Numerical values of the coefficients $C_1$\ldots$C_4$ are listed in
table~\ref{tab:coefficients}.

\begin{table}[h]
\caption{Coefficients  $\mathbf C_1$\ldots$\mathbf C_4$ describing the
  contributions of cilia beyond $\mathbf o$ periods, for different orders
  $\mathbf o$.}
\label{tab:coefficients}
\begin{tabular}{lllll}
\hline
$o$ & $C_1$ & $C_2$ & $C_3$ & $C_4$\\
\hline
0 & 4.51681&14.0613&-2.60821&-12.8518 \\
1 & 1.80961& 0.680011& 0.182076& -0.210573\\
2 & 1.1135& 0.161671& 0.0474549& -0.0445071\\
3 & 0.801452&0.0608975& 0.0181854& -0.0163511\\
\hline
\end{tabular}
\end{table}

With discretized time steps the expression for velocity
(\ref{eq:mobilityarray1}) becomes
\begin{equation}
\begin{split}
  \label{eq:mobilityarrayd}
   \dot{\ve x}_{i} (\tau)&= \sum_{j=1}^{N}\sum_{\alpha,\beta} M'_{i,j;\alpha,
     \beta} (\tau) \ve F_j(\tau- \alpha N_f\kappa_x -\beta N_f
   \kappa_y) \\ &=\sum_{j=1}^{N}\sum_{\tau'=1}^{N_S}\tilde M_{i,j;
     \tau, \tau'}  \ve F_j(\tau')
 \end{split}
\end{equation}
with $ \tilde M_{i,j;\tau, \tau'}=\sum_{\alpha,\beta} M'_{i,j;\alpha,
  \beta}(\tau) [\tau'\equiv \tau-\alpha N_f\kappa_x -\beta N_f \kappa_y
\pmod{N_S}]$ denoting the generalized mobility matrix that now couples the
forces and velocities at different time steps. The brackets $[]$ denote a
function which is $1$ if the condition is fulfilled and $0$ otherwise. We also
set the number of time steps to be a multiple of the lattice size, $N_S=N_f
N_a$. $\tilde M$ is in total a $3N N_S\times 3 N N_S$ dimensional matrix,
which can be re-ordered into a block-diagonal form with $N_f$ blocks for
greater numerical efficiency.

We calculate corresponding generalized friction matrix by inverting the
generalized mobility matrix,
\begin{equation}
  \label{eq:genfriction}
  \tilde \Gamma = \tilde M^{-1}\;.
\end{equation}
The friction matrix now allows us to calculate the forces if we know the
velocities of all beads at all times,
\begin{equation}
{\ve F_i}(\tau)=\sum_{j=1}^{N} \sum_{\tau'=1}^{N_S}  \tilde \Gamma_{i,j; \tau ,\tau'}
\dot{\ve x}_j(\tau')\;.
\end{equation}
Within the finite difference approximation this leads to equations 
\begin{equation}
\begin{split}
Q=\frac 1 {\pi \eta N_S} \sum_{\tau,\tau'=1}^{N_S}\sum_{i,j=1}^{N} \frac{z_{i}(
  \tau+1)\Gamma_{i,j;\tau+1,\tau'+1}+z_{i}( \tau)
\Gamma_{i,j;\tau,\tau'}}{2} \,\\ \times  \frac{{\bf x}_j(\tau'+1)-{\bf
  x}_j(\tau')}{\Delta t}
\end{split}
\end{equation}
\begin{equation}
\begin{split}
  P=\frac 1 {N_S}\sum_{\tau,\tau'=1}^{N_S}&\sum_{i,j=1}^{N}  \frac{\Gamma_{i,j;\tau+1,\tau'+1}+
    \Gamma_{i,j;\tau,\tau'}}{2}   \,\\ &\times \frac{\big({\bf x}_i(\tau'+1)-{\bf x}_i(\tau') \big) \big ({\bf x}_j(\tau'+1)-{\bf x}_j(\tau') \big)}{(\Delta t)^2}.
\end{split}
\end{equation}
Now we can numerically optimize the quantity $Q/\sqrt{P}$ as a function of
the angles that parameterize the beat shape (in total $2(N-1) N_S$
variables). We initialize the system in a way that only solutions with
clockwise rotation (as seen from above) are considered. The constraints are
similar to those on a single cilium, except that hard-core repulsion between
neighboring cilia has to be taken into account, too.

In conclusion, the combination of an efficient description of hydrodynamics,
the usage of periodic boundary conditions and the translational invariance of
the metachronal wave allowed us to reduce the optimization problem to one with
the same number of variables as for a single cilium. The computational demand
to calculate the efficiency for a certain set of variables scales as
$((2o+1)N_a)^2$, rather than the fourth power which would be the case if all
cilia were treated independently. The optimization for one wave vector
typically takes a few days on one core of the Intel Xeon E5520 CPU. To find
the global maximum we have to repeat the calculation with all wave vectors,
$N_a^2$ in total.

\bibliography{bib/cilia,bib/biomot}

\begin{thebibliography}{32}
\expandafter\ifx\csname natexlab\endcsname\relax\def\natexlab#1{#1}\fi
\expandafter\ifx\csname bibnamefont\endcsname\relax
  \def\bibnamefont#1{#1}\fi
\expandafter\ifx\csname bibfnamefont\endcsname\relax
  \def\bibfnamefont#1{#1}\fi
\expandafter\ifx\csname citenamefont\endcsname\relax
  \def\citenamefont#1{#1}\fi
\expandafter\ifx\csname url\endcsname\relax
  \def\url#1{\texttt{#1}}\fi
\expandafter\ifx\csname urlprefix\endcsname\relax\def\urlprefix{URL }\fi
\providecommand{\bibinfo}[2]{#2}
\providecommand{\eprint}[2][]{\url{#2}}

\bibitem[{\citenamefont{Kushmerick and Davies}(1969)}]{Kushmerick.Davies1969}
\bibinfo{author}{\bibfnamefont{M.~J.} \bibnamefont{Kushmerick}}
  \bibnamefont{and} \bibinfo{author}{\bibfnamefont{R.~E.}
  \bibnamefont{Davies}}, \bibinfo{journal}{Proc.~R.~Soc.~Lond.~B Biol.~Sci.}
  \textbf{\bibinfo{volume}{174}}, \bibinfo{pages}{315} (\bibinfo{year}{1969}).

\bibitem[{\citenamefont{Yoshida et~al.}(2001)\citenamefont{Yoshida, Muneyuki,
  and Hisabori}}]{Yoshida.Hisabori2001}
\bibinfo{author}{\bibfnamefont{M.}~\bibnamefont{Yoshida}},
  \bibinfo{author}{\bibfnamefont{E.}~\bibnamefont{Muneyuki}}, \bibnamefont{and}
  \bibinfo{author}{\bibfnamefont{T.}~\bibnamefont{Hisabori}},
  \bibinfo{journal}{Nat.\ Rev.\ Mol.\ Cell.\ Biol.}
  \textbf{\bibinfo{volume}{2}}, \bibinfo{pages}{669} (\bibinfo{year}{2001}).

\bibitem[{\citenamefont{Sleigh}(1974)}]{Sleigh74}
\bibinfo{editor}{\bibfnamefont{M.~A.} \bibnamefont{Sleigh}}, ed.,
  \emph{\bibinfo{title}{Cilia and Flagella}} (\bibinfo{publisher}{Academic
  Press}, \bibinfo{address}{London}, \bibinfo{year}{1974}).

\bibitem[{\citenamefont{Lighthill}(1952)}]{Lighthill1952}
\bibinfo{author}{\bibfnamefont{M.~J.} \bibnamefont{Lighthill}},
  \bibinfo{journal}{Comm.\ Pure Appl.\ Math.} \textbf{\bibinfo{volume}{5}},
  \bibinfo{pages}{109} (\bibinfo{year}{1952}).

\bibitem[{\citenamefont{Michelin and Lauga}(2010)}]{Michelin.Lauga2010}
\bibinfo{author}{\bibfnamefont{S.}~\bibnamefont{Michelin}} \bibnamefont{and}
  \bibinfo{author}{\bibfnamefont{E.}~\bibnamefont{Lauga}},
  \bibinfo{journal}{Phys.\ Fluids} \textbf{\bibinfo{volume}{22}},
  \bibinfo{pages}{111901} (\bibinfo{year}{2010}).

\bibitem[{\citenamefont{Purcell}(1997)}]{Purcell1997}
\bibinfo{author}{\bibfnamefont{E.~M.} \bibnamefont{Purcell}},
  \bibinfo{journal}{Proc.\ Natl.\ Acad.\ Sci.\ USA}
  \textbf{\bibinfo{volume}{94}}, \bibinfo{pages}{11307} (\bibinfo{year}{1997}).

\bibitem[{\citenamefont{Chattopadhyay et~al.}(2006)\citenamefont{Chattopadhyay,
  Moldovan, Yeung, and Wu}}]{Chattopadhyay.Wu2006}
\bibinfo{author}{\bibfnamefont{S.}~\bibnamefont{Chattopadhyay}},
  \bibinfo{author}{\bibfnamefont{R.}~\bibnamefont{Moldovan}},
  \bibinfo{author}{\bibfnamefont{C.}~\bibnamefont{Yeung}}, \bibnamefont{and}
  \bibinfo{author}{\bibfnamefont{X.~L.} \bibnamefont{Wu}},
  \bibinfo{journal}{Proc.\ Natl.\ Acad.\ Sci.\ USA}
  \textbf{\bibinfo{volume}{103}}, \bibinfo{pages}{13712}
  (\bibinfo{year}{2006}).

\bibitem[{\citenamefont{Purcell}(1977)}]{Purcell1977}
\bibinfo{author}{\bibfnamefont{E.~M.} \bibnamefont{Purcell}},
  \bibinfo{journal}{Am.\ J.\ Phys.} \textbf{\bibinfo{volume}{45}},
  \bibinfo{pages}{3} (\bibinfo{year}{1977}).

\bibitem[{\citenamefont{Katsu-Kimura et~al.}(2009)\citenamefont{Katsu-Kimura,
  Nakaya, Baba, and Mogami}}]{Katsu-Kimura.Mogami2009}
\bibinfo{author}{\bibfnamefont{Y.}~\bibnamefont{Katsu-Kimura}},
  \bibinfo{author}{\bibfnamefont{F.}~\bibnamefont{Nakaya}},
  \bibinfo{author}{\bibfnamefont{S.~A.} \bibnamefont{Baba}}, \bibnamefont{and}
  \bibinfo{author}{\bibfnamefont{Y.}~\bibnamefont{Mogami}},
  \bibinfo{journal}{J. Exp.\ Biol.} \textbf{\bibinfo{volume}{212}},
  \bibinfo{pages}{1819} (\bibinfo{year}{2009}).

\bibitem[{\citenamefont{Supatto et~al.}(2008)\citenamefont{Supatto, Fraser, and
  Vermot}}]{Supatto.Vermot2008}
\bibinfo{author}{\bibfnamefont{W.}~\bibnamefont{Supatto}},
  \bibinfo{author}{\bibfnamefont{S.~E.} \bibnamefont{Fraser}},
  \bibnamefont{and} \bibinfo{author}{\bibfnamefont{J.}~\bibnamefont{Vermot}},
  \bibinfo{journal}{Biophys.~J.} \textbf{\bibinfo{volume}{95}},
  \bibinfo{pages}{L29} (\bibinfo{year}{2008}).

\bibitem[{\citenamefont{Tam and Hosoi}(2007)}]{Tam.Hosoi2007}
\bibinfo{author}{\bibfnamefont{D.}~\bibnamefont{Tam}} \bibnamefont{and}
  \bibinfo{author}{\bibfnamefont{A.~E.} \bibnamefont{Hosoi}},
  \bibinfo{journal}{Phys.~Rev.~Lett.} \textbf{\bibinfo{volume}{98}},
  \bibinfo{pages}{068105} (\bibinfo{year}{2007}).

\bibitem[{\citenamefont{Spagnolie and Lauga}(2010)}]{Spagnolie.Lauga2010}
\bibinfo{author}{\bibfnamefont{S.~E.} \bibnamefont{Spagnolie}}
  \bibnamefont{and} \bibinfo{author}{\bibfnamefont{E.}~\bibnamefont{Lauga}},
  \bibinfo{journal}{Phys.\ Fluids} \textbf{\bibinfo{volume}{22}},
  \bibinfo{pages}{031901} (\bibinfo{year}{2010}).

\bibitem[{\citenamefont{Avron et~al.}(2004)\citenamefont{Avron, Gat, and
  Kenneth}}]{Avron.Kenneth2004}
\bibinfo{author}{\bibfnamefont{J.~E.} \bibnamefont{Avron}},
  \bibinfo{author}{\bibfnamefont{O.}~\bibnamefont{Gat}}, \bibnamefont{and}
  \bibinfo{author}{\bibfnamefont{O.}~\bibnamefont{Kenneth}},
  \bibinfo{journal}{Phys.~Rev.~Lett.} \textbf{\bibinfo{volume}{93}},
  \bibinfo{pages}{186001} (\bibinfo{year}{2004}).

\bibitem[{\citenamefont{Alouges et~al.}(2009)\citenamefont{Alouges, DeSimone,
  and Lefebvre}}]{Alouges.Lefebvre2009}
\bibinfo{author}{\bibfnamefont{F.}~\bibnamefont{Alouges}},
  \bibinfo{author}{\bibfnamefont{A.}~\bibnamefont{DeSimone}}, \bibnamefont{and}
  \bibinfo{author}{\bibfnamefont{A.}~\bibnamefont{Lefebvre}},
  \bibinfo{journal}{Eur.\ Phys.\ J. E Soft Matter}
  \textbf{\bibinfo{volume}{28}}, \bibinfo{pages}{279} (\bibinfo{year}{2009}).

\bibitem[{\citenamefont{Tam and Hosoi}(2011)}]{Tam.Hosoi2011}
\bibinfo{author}{\bibfnamefont{D.}~\bibnamefont{Tam}} \bibnamefont{and}
  \bibinfo{author}{\bibfnamefont{A.~E.} \bibnamefont{Hosoi}},
  \bibinfo{journal}{Proc.\ Natl.\ Acad.\ Sci.\ USA}
  \textbf{\bibinfo{volume}{108}}, \bibinfo{pages}{1001} (\bibinfo{year}{2011}).

\bibitem[{\citenamefont{Smith et~al.}(2008)\citenamefont{Smith, Blake, and
  Gaffney}}]{Smith.Gaffney2008}
\bibinfo{author}{\bibfnamefont{D.~J.} \bibnamefont{Smith}},
  \bibinfo{author}{\bibfnamefont{J.~R.} \bibnamefont{Blake}}, \bibnamefont{and}
  \bibinfo{author}{\bibfnamefont{E.~A.} \bibnamefont{Gaffney}},
  \bibinfo{journal}{J. R. Soc.\ Interface} \textbf{\bibinfo{volume}{5}},
  \bibinfo{pages}{567} (\bibinfo{year}{2008}).

\bibitem[{\citenamefont{Vilfan and J{\" u}licher}(2006)}]{vilfan2006a}
\bibinfo{author}{\bibfnamefont{A.}~\bibnamefont{Vilfan}} \bibnamefont{and}
  \bibinfo{author}{\bibfnamefont{F.}~\bibnamefont{J{\" u}licher}},
  \bibinfo{journal}{Phys.~Rev.~Lett.} \textbf{\bibinfo{volume}{96}},
  \bibinfo{pages}{058102} (\bibinfo{year}{2006}).

\bibitem[{\citenamefont{Jülicher and Prost}(2009)}]{Julicher.Prost2009}
\bibinfo{author}{\bibfnamefont{F.}~\bibnamefont{Jülicher}} \bibnamefont{and}
  \bibinfo{author}{\bibfnamefont{J.}~\bibnamefont{Prost}},
  \bibinfo{journal}{Eur.\ Phys.\ J. E Soft Matter}
  \textbf{\bibinfo{volume}{29}}, \bibinfo{pages}{27} (\bibinfo{year}{2009}).

\bibitem[{\citenamefont{Gueron and
  Levit-Gurevich}(1999)}]{Gueron.Levit-Gurevich1999}
\bibinfo{author}{\bibfnamefont{S.}~\bibnamefont{Gueron}} \bibnamefont{and}
  \bibinfo{author}{\bibfnamefont{K.}~\bibnamefont{Levit-Gurevich}},
  \bibinfo{journal}{Proc.\ Natl.\ Acad.\ Sci.\ USA}
  \textbf{\bibinfo{volume}{96}}, \bibinfo{pages}{12240} (\bibinfo{year}{1999}).

\bibitem[{\citenamefont{Gauger et~al.}(2009)\citenamefont{Gauger, Downton, and
  Stark}}]{Gauger.Stark2009}
\bibinfo{author}{\bibfnamefont{E.~M.} \bibnamefont{Gauger}},
  \bibinfo{author}{\bibfnamefont{M.~T.} \bibnamefont{Downton}},
  \bibnamefont{and} \bibinfo{author}{\bibfnamefont{H.}~\bibnamefont{Stark}},
  \bibinfo{journal}{Eur.\ Phys.\ J. E Soft Matter}
  \textbf{\bibinfo{volume}{28}}, \bibinfo{pages}{231} (\bibinfo{year}{2009}).

\bibitem[{\citenamefont{Happel and Brenner}(1983)}]{Happel.Brenner}
\bibinfo{author}{\bibfnamefont{J.}~\bibnamefont{Happel}} \bibnamefont{and}
  \bibinfo{author}{\bibfnamefont{H.}~\bibnamefont{Brenner}},
  \emph{\bibinfo{title}{Low Reynolds Number Hydrodynamics}}
  (\bibinfo{publisher}{Kluwer, Dodrecht}, \bibinfo{year}{1983}).

\bibitem[{\citenamefont{Blake}(1971)}]{Blake.1971}
\bibinfo{author}{\bibfnamefont{J.~R.} \bibnamefont{Blake}},
  \bibinfo{journal}{Proc.~Camb.~Phil.~Soc.} \textbf{\bibinfo{volume}{70}},
  \bibinfo{pages}{303} (\bibinfo{year}{1971}).

\bibitem[{\citenamefont{Vilfan et~al.}(2010)\citenamefont{Vilfan, Potočnik,
  Kavčič, Osterman, Poberaj, Vilfan, and Babič}}]{Vilfan.Babic2010}
\bibinfo{author}{\bibfnamefont{M.}~\bibnamefont{Vilfan}},
  \bibinfo{author}{\bibfnamefont{A.}~\bibnamefont{Potočnik}},
  \bibinfo{author}{\bibfnamefont{B.}~\bibnamefont{Kavčič}},
  \bibinfo{author}{\bibfnamefont{N.}~\bibnamefont{Osterman}},
  \bibinfo{author}{\bibfnamefont{I.}~\bibnamefont{Poberaj}},
  \bibinfo{author}{\bibfnamefont{A.}~\bibnamefont{Vilfan}}, \bibnamefont{and}
  \bibinfo{author}{\bibfnamefont{D.}~\bibnamefont{Babič}},
  \bibinfo{journal}{Proc.\ Natl.\ Acad.\ Sci.\ USA}
  \textbf{\bibinfo{volume}{107}}, \bibinfo{pages}{1844} (\bibinfo{year}{2010}).

\bibitem[{\citenamefont{Najafi and
  Golestanian}(2004)}]{Najafi.Golestanian.2004}
\bibinfo{author}{\bibfnamefont{A.}~\bibnamefont{Najafi}} \bibnamefont{and}
  \bibinfo{author}{\bibfnamefont{R.}~\bibnamefont{Golestanian}},
  \bibinfo{journal}{Phys.~Rev.~E} \textbf{\bibinfo{volume}{69}},
  \bibinfo{pages}{062901} (\bibinfo{year}{2004}).

\bibitem[{\citenamefont{Downton and Stark}(2009)}]{Downton.Stark2009}
\bibinfo{author}{\bibfnamefont{M.~T.} \bibnamefont{Downton}} \bibnamefont{and}
  \bibinfo{author}{\bibfnamefont{H.}~\bibnamefont{Stark}},
  \bibinfo{journal}{Europhys.~Lett.} \textbf{\bibinfo{volume}{85}},
  \bibinfo{pages}{44002} (\bibinfo{year}{2009}).

\bibitem[{\citenamefont{Stone and Samuel}(1996)}]{Stone.Samuel1996}
\bibinfo{author}{\bibfnamefont{H.~A.} \bibnamefont{Stone}} \bibnamefont{and}
  \bibinfo{author}{\bibfnamefont{A.~D.} \bibnamefont{Samuel}},
  \bibinfo{journal}{Phys.~Rev.~Lett.} \textbf{\bibinfo{volume}{77}},
  \bibinfo{pages}{4102} (\bibinfo{year}{1996}).

\bibitem[{\citenamefont{Brennen and Winet}(1977)}]{Brennen.Winet1977}
\bibinfo{author}{\bibfnamefont{C.}~\bibnamefont{Brennen}} \bibnamefont{and}
  \bibinfo{author}{\bibfnamefont{H.}~\bibnamefont{Winet}},
  \bibinfo{journal}{Ann. Rev. Fluid Mech.} \textbf{\bibinfo{volume}{9}},
  \bibinfo{pages}{339} (\bibinfo{year}{1977}).

\bibitem[{\citenamefont{Gueron et~al.}(1997)\citenamefont{Gueron,
  Levit-Gurevich, Liron, and Blum}}]{Gueron.Blum1997}
\bibinfo{author}{\bibfnamefont{S.}~\bibnamefont{Gueron}},
  \bibinfo{author}{\bibfnamefont{K.}~\bibnamefont{Levit-Gurevich}},
  \bibinfo{author}{\bibfnamefont{N.}~\bibnamefont{Liron}}, \bibnamefont{and}
  \bibinfo{author}{\bibfnamefont{J.~J.} \bibnamefont{Blum}},
  \bibinfo{journal}{Proc.\ Natl.\ Acad.\ Sci.\ USA}
  \textbf{\bibinfo{volume}{94}}, \bibinfo{pages}{6001} (\bibinfo{year}{1997}).

\bibitem[{\citenamefont{Sleigh}(1969)}]{Sleigh1969}
\bibinfo{author}{\bibfnamefont{M.~A.} \bibnamefont{Sleigh}},
  \bibinfo{journal}{Int.\ Rev.\ Cytol.} \textbf{\bibinfo{volume}{25}},
  \bibinfo{pages}{31} (\bibinfo{year}{1969}).

\bibitem[{\citenamefont{Machemer}(1972)}]{Machemer1972}
\bibinfo{author}{\bibfnamefont{H.}~\bibnamefont{Machemer}},
  \bibinfo{journal}{J. Exp.\ Biol.} \textbf{\bibinfo{volume}{57}},
  \bibinfo{pages}{239} (\bibinfo{year}{1972}).

\bibitem[{\citenamefont{Gheber and Priel}(1990)}]{Gheber.Priel1990a}
\bibinfo{author}{\bibfnamefont{L.}~\bibnamefont{Gheber}} \bibnamefont{and}
  \bibinfo{author}{\bibfnamefont{Z.}~\bibnamefont{Priel}},
  \bibinfo{journal}{Cell~Motil.~Cytoskeleton} \textbf{\bibinfo{volume}{16}},
  \bibinfo{pages}{167} (\bibinfo{year}{1990}).

\bibitem[{\citenamefont{Sleigh and Blake}(1977)}]{Sleigh.Blake1977}
\bibinfo{author}{\bibfnamefont{M.~A.} \bibnamefont{Sleigh}} \bibnamefont{and}
  \bibinfo{author}{\bibfnamefont{J.~R.} \bibnamefont{Blake}}, in
  \emph{\bibinfo{booktitle}{Scale Effects in Animal Locomotion}}, edited by
  \bibinfo{editor}{\bibfnamefont{T.~J.} \bibnamefont{Pedley}}
  (\bibinfo{publisher}{Academic Press}, \bibinfo{address}{London},
  \bibinfo{year}{1977}), pp. \bibinfo{pages}{243--256}.

\end{thebibliography}

\end{document}